\documentclass[osajnl,twocolumn,showpacs,superscriptaddress,11pt]{revtex4-1} %% use 11pt for Applied Optics
\usepackage{amsmath,amssymb,graphicx,xcolor}
\begin{document}

\title{Phase retrieval of reflection and transmission coefficients\\[3mm] 
from Kramers-Kronig relations}

\author{Boris Gralak}
\affiliation{CNRS, Aix-Marseille Universit\'e, Ecole Centrale 
Marseille, Institut Fresnel, 13397 Marseille cedex 20, France}

\author{Michel Lequime}
\affiliation{CNRS, Aix-Marseille Universit\'e, Ecole Centrale 
Marseille, Institut Fresnel, 13397 Marseille cedex 20, France}

\author{Myriam Zerrad}
\affiliation{CNRS, Aix-Marseille Universit\'e, Ecole Centrale 
Marseille, Institut Fresnel, 13397 Marseille cedex 20, France}

\author{Claude Amra}
\affiliation{CNRS, Aix-Marseille Universit\'e, Ecole Centrale 
Marseille, Institut Fresnel, 13397 Marseille cedex 20, France}

\begin{abstract}
Analytic and passivity properties of reflection and transmission coefficients of thin-film 
multilayered stacks are investigated. Using a rigorous formalism based on the inverse Helmholtz 
operator, properties associated to causality principle and passivity are established when both 
temporal frequency and spatial wavevector are continued in the complex plane. This result 
extends the range of situations where the Kramers-Kronig relations can be 
used to deduce the phase from the intensity. In particular, it is rigorously shown that 
Kramers-Kronig relations for reflection and transmission coefficients remain valid at a fixed angle 
of incidence. Possibilities to exploit the new relationships are discussed. 
\end{abstract}

%\ocis{(310.6188) Spectral properties; (310.6860) Thin films, optical properties; 
%(260.2030) Dispersion; (230.4170) Multilayers.}% REPLACE WITH CORRECT OCIS CODES FOR YOUR ARTICLE
                          % NOTE: \ocis{} IS ALIASED TO \pacs{} BUT MUST
                          % FORMAT THE TERMS CORRECTLY FOR EACH JOURNAL

\maketitle %% required

%--------------------------------------------------------------------------------------------------
% notations
%--------------------------------------------------------------------------------------------------
\def\x{\text{\bfseries\sffamily\textit{x}}}
\def\xp{\text{\bfseries\sffamily\textit{x}}_{\!\text{\sffamily\textit{p}}}}
\def\xp{\text{\bfseries\sffamily\textit{v}}}
\def\xs{\text{\sffamily\textit{x}}}
\def\ds{\text{\sffamily\textit{d}}}
\def\y{\text{\bfseries\sffamily\textit{y}}}
\def\ys{\text{\sffamily\textit{y}}}
\def\k{\text{\bfseries\sffamily\textit{k}}}
\def\ks{\text{\sffamily\textit{k}}}
\def\kp{\text{\bfseries\sffamily\textit{k}}_{\!\text{\sffamily\textit{p}}}}
\def\kp{\text{\bfseries\sffamily\textit{k}}}
\def\kp{\boldsymbol{\kappa}}
\def\kp{\boldsymbol{\k}}
\def\u{\text{\bfseries\sffamily\textit{u}}}
\def\E{\text{\bfseries\sffamily\textit{E}}}
\def\P{\text{\bfseries\sffamily\textit{P}}}
\def\B{\text{\bfseries\sffamily\textit{B}}}
\def\D{\text{\bfseries\sffamily\textit{D}}}
\def\A{\text{\bfseries\sffamily\textit{A}}}
\def\J{\text{\bfseries\sffamily\textit{J}}}
\def\F{\text{\sffamily\textit{F}}}
\def\t{\text{\sffamily\textit{t}}}
\def\s{\text{\sffamily\textit{s}}}
\def\nui{\text{\sffamily\textit{$\nu$}}}
\def\f{\text{\sffamily{f}}}
\def\f{\text{\bfseries\sffamily\textit{f}}}
\def\c{\text{\sffamily\textit{c}}}
\def\rot{\boldsymbol{\nabla \times }}
\def\inv{\mathsf{R}}
\def\dt{\partial_{\t}}
\def\hatLE{\hspace*{0.35mm}\widehat{\hspace*{-0.35mm}\text{\bfseries\sffamily\textit{E}}\hspace*{0.35mm}}\hspace*{-0.35mm}}
\def\LJ{\,\widehat{\!\text{\bfseries\sffamily\textit{J}}\,}\!}
\def\LD{\,\widehat{\!\text{\bfseries\sffamily\textit{D}}\,}\!}
% Autre notation pour les champs
\def\E{\text{\bfseries\textit{E}}}
\def\P{\text{\bfseries\textit{P}}}
\def\B{\text{\bfseries\textit{B}}}
\def\D{\text{\bfseries\textit{D}}}
\def\A{\text{\bfseries\textit{A}}}
\def\J{\text{\bfseries\textit{J}}}
\def\F{\text{\textit{F}}}
% et autre notation pour les transformées de Laplace
\def\LE{\text{\bfseries\sffamily\textit{E}}}
\def\LP{\text{\bfseries\sffamily\textit{P}}}
\def\LB{\text{\bfseries\sffamily\textit{B}}}
\def\LD{\text{\bfseries\sffamily\textit{D}}}
\def\LA{\text{\bfseries\sffamily\textit{A}}}
\def\LJ{\text{\bfseries\sffamily\textit{J}}}
\def\LF{\text{\sffamily\textit{F}}}
\def\ep{\varepsilon}
\def\om{\omega}
\def\resub{\color{black}}
%--------------------------------------------------------------------------------------------------
\section{Introduction}
%--------------------------------------------------------------------------------------------------

Phase data have become a key in multilayer optics since they drive resonance effects and broad-band 
properties in most optical coatings \cite{Macleod,Baum,Thelen,Tikhon}, 
with additional recent applications in the field of chirped 
mirrors \cite{Pervak}. Moreover such data provide a complementary characterization tool to probe multilayers 
and solve inverse problems, in order to check the agreement with the expected design, etc...
For these reasons a number of optical techniques were developed to investigate phase data, in addition 
to reflection and transmission energy coefficients. Ellipsometry has widely been used in this 
context, but provides differential phase data not available at normal illumination. Absolute phase data 
are more complex to extract and often involve interferential techniques \cite{Pervak09,Xue09} 
more sensitive to the surrounding.
Within this framework complementary techniques to extract phase data at low-cost with high accuracy 
remains a challenge for a number of applications including optical microscopy. Among them, the 
Kramers-Kronig techniques deserve to be furthermore explored.

Kramers-Kronig relationships are classically based on a causality principle which describes the 
temporal behavior of a material submitted to excitation \cite{Jackson,Landau8}. Such behavior 
%{\bf{of induced polarization in materials}} 
can be seen as the result 
of a linear filter, that is, the result of a convolution product between the input excitation and 
another function characteristic of the material microstructure (permittivity, permeability…); this 
last function vanishes at negative instants, so that the material {\resub{response}} at time $t$ depends 
on all excitation values at lower instants ($\t' < \t$). Due to this intuitive property, 
mathematical transformations emphasize specific integrals in the Fourier plane that connect 
the real (Re$\, \varepsilon$) and imaginary (Im$\,\varepsilon$) parts of permittivity (for instance)
\cite{Jackson,Landau8}. In other words, it is well established that Im$\,\varepsilon$ at one 
frequency can be deduced from the values of Re$\,\varepsilon$ at all other frequencies, and conversely.
This is a general result for signals so-called causal signals. Hence the amplitude reflection 
from a multilayer should also follow the Kramers-Kronig criterion, since the reflected field can 
also be written as the result of a linear filter, where the characteristic function is a double 
inverse Fourier transform (over temporal and spatial frequency) of the reflection coefficient. 
In this case the consequence is that the spectral phase of the stack can be retrieved from the 
intensity spectral properties of the same stack. Several authors have worked on 
this topic in various situations of multilayers \cite{GO91,TBP97}.
However, as pointed out in the literature, serious difficulties 
remain and result from the existence of zeros in reflection spectra
{\resub{since they produce branch points and cuts when the complex logarithm 
is used \cite{Nasha95,Lee97,Andre10} to separate the phase from the modulus.}}
In addition, to our knowledge, a rigorous proof of the possibility 
to use Kramers-Kronig relation at oblique incidence has not been established. 

In this paper, we use the techniques derived in \cite{GT10} to show new properties 
of reflection and transmission coefficients of multilayered stacks. First, the usual analytic 
properties with respect to the complex frequency $z$, currently stated at normal incidence 
\cite{Tikhon93, TBP97}, are generalized to the complex wavevector $\kp$. This new property 
extends the possibility to use the analyticity for all angles of incidence {\resub{from 0 to 90 degrees}}. 
Next, a second property, denominated by ``passivity property'', shows that the sum of the 
reflection coefficient with a phase shift, $\exp[-i\beta_\circ(\om,\kp)\ds] + r(\omega,\kp)$, 
cannot vanish in an appropriate domain of the complex frequency and wavevector. This passivity 
property makes it possible to apply the complex logarithm without alteration of the analytic 
properties with respect to the complex frequency and 
wavevector. They are used to propose alternative solutions to retrieve the phase of reflection and 
transmission coefficients. In particular, since the quantity 
{\resub{$\{ 1+\exp[i\beta_\circ(\om,\kp)\ds] \, r(\omega,\kp) \}$}} cannot vanish, 
the proposed alternative solutions for the reflection coefficient do not use Blaschke factors 
\cite{Tikhon93, TBP97}.

%--------------------------------------------------------------------------------------------------
\section{Background}
%--------------------------------------------------------------------------------------------------

%In order to address the most general properties, the inverse of the Helmholtz operator is directly 
%considered. Then many usual quantities can be deduced : the usual inverse Helmholtz operator, the 
%Green's function, $R(z,\kp)$ and $T(z,\kp)$\dots

\begin{figure}
\centerline{\includegraphics[width=.8\columnwidth]{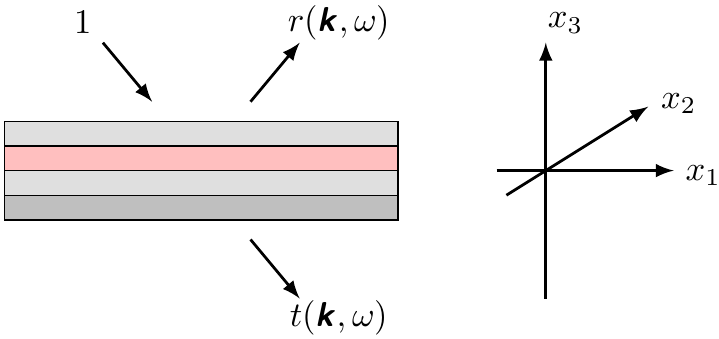}}
\caption{Reflection and transmission coefficients of a thin film multilayered stack.\label{fig1}}
\end{figure}

Throughout this article an orthonormal basis is used: every vector $\x$ in $\mathbb{R}^3$ 
is described by its three components $\xs_1$, $\xs_2$ and $\xs_3$ (see Fig.\ \ref{fig1}).
We start with the usual Helmholtz equation in non magnetic, isotropic and linear media, 
and in the absence of sources. The time-harmonic electric field $\LE(\x,\om)$ 
is the solution of
\begin{equation}
\om^2 \mu_\circ \ep(\x,\om) \LE(\x,\om) - \rot \rot \LE(\x,\om) = 0 \, .
\label{Helm1}
\end{equation}
where $\rot$ is the curl operator, $\om$ is the temporal pulsation resulting from the Fourier 
decomposition with respect to time, $\mu_\circ$ is the vacuum permeability,
and $\ep(\x,\om)$ the frequency-dependent permittivity. 

In the case of a stack of homogeneous layers invariant in $\xs_1$ and $\xs_2$ 
directions (see Fig.\ \ref{fig1}), a Fourier decomposition 
$\LE(\x,\om) \longrightarrow \hatLE (\kp,\xs_3,\om)$ with respect to the 
space variables $(\xs_1,\xs_2) = \xp$ is performed,
\begin{equation}
\hatLE (\kp,\xs_3,\om) = \dfrac{1}{4 \pi^2} 
\displaystyle\int_{\mathbb{R}^2} \exp[-i \kp \cdot \xp ] 
\LE(\xp,\xs_3,\om) \, d\xp \, ,
\label{Fx1x2}
\end{equation}
where $\kp = (\ks_1,\ks_2)$ is the two-dimensional wave vector associated with the 
invariance planes of the geometry. Then, the electric field $\hatLE (\kp,\xs_3,\om)$ 
can be determined in all the multilayered stack and the surrounding homogeneous media 
using the admittance formalism \cite{Macleod} or transfer matrix \cite{Macleod} 
formalisms. In particular, the field 
$\hatLE (\kp,\xs_3,\om)$ is expressed in the surrounding homogeneous media in terms of 
the reflection and transmission coefficients $r(\kp, \om)$ and $t(\kp, \om)$ of the field 
amplitude.

In this paper, it is proposed to study the properties of the reflection and 
transmission coefficients with respect to both the temporal frequency and spatial wave vector 
extended to the complex plane: in particular, the frequency becomes the complex number 
$z = \om + i \eta$, where $\eta$ is the positive imaginary part.
%Also, the analytic properties with respect 
%to the complex components of the wave vector $\kp$ will be approached. 
It is stressed that  
these properties are closely connected, not to say equivalent, to causality principle. 
Indeed, from the Paley-Wiener theorem (see theorem IX.11 in \cite{RS2}), analytic properties of a function 
imposes to its Fourier transform to vanish in a domain of the Fourier space. For exemple, 
the permittivity $\ep(\x,z)$ is an analytic function in the upper half plane of complex 
frequencies $z = \om + i \eta$ with positive imaginary part $\eta$. Using that $\ep(\x,z)$ 
tends to the vacuum permittivity $\ep_\circ$ when $|z| \rightarrow \infty$, it follows that 
the electric susceptibility defined by 
\begin{equation} 
\chi(\x,\t) = \int_{\mathbb{R}} \exp[- i \om \t ] \, [\ep(\x,\om) / \ep_\circ - 1] \, d\om \, 
\label{chi}
\end{equation}
vanishes for negative values of the time variable $\t$ [here, notice that the symbol 
``$\t$'' is used for the time variable, while the transmission coefficient is denoted by 
``$t(\kp, \om)$'']. Indeed, in that case, the integral 
over the frequency in the equation above, can be computed by closing the integration path 
by a semi circle in upper half plane which, in combination with the Cauchy's theorem, yields 
$\chi(\x,\t) = 0$ if $\t<0$.  

The amplitudes $r(\kp, \om)$ and $t(\kp, \om)$ of reflected and transmitted waves are 
generally used to compute the Green's function of the multilayered stack \cite{Tom95}. 
Here, on the contrary, it is proposed to use the knowledge of the Green's function to deduce 
the properties of the reflection and transmission coefficients. In practice, these properties 
will be derived from the inverse of the Helmholtz operator (\ref{Helm1}) whose the Green's 
function is nothing else than the kernel. 
In order to address the most general properties of reflection and transmission coefficients, 
the Helmholtz operator is rigorously defined in the next section using the auxiliary field 
formalism \cite{Tip98}. 

%--------------------------------------------------------------------------------------------------
\section{Inverse of Helmholtz operator}
%--------------------------------------------------------------------------------------------------

We start with Maxwell's equations. Let $\E(\x,\t)$, $\B(\x,\t)$ and $\P(\x,\t)$ be respectively 
the time-dependent electric, magnetic and polarization fields. Then,
\begin{equation}
\begin{array}{l}
\ep_\circ \, \dt \E(\x,\t) + \dt \P(\x,\t) = \rot \B(\x,\t) / \mu_\circ \, , \\[2mm]
\dt \B(\x,\t) = - \rot\E(\x,\t) \, , 
\end{array}
\label{Max-eq}
\end{equation}
where $\dt$ is the partial derivative with respect to time. 
In addition to these equations, the electric field is related to the polarization through the 
constitutive equation
\begin{equation}
\P(\x,\t) = \ep_\circ \displaystyle\int_{-\infty}^{\t} \chi(\x,\t-\s) \E(\x,\s) \, d \s \, , 
\label{const-eq}
\end{equation}
where $\chi(\x,\t)$ is the electric susceptibility that vanishes for negative times: 
$\chi(\x,\t) = 0$ if $\t<0$. According to (\ref{chi}), the dielectric permittivity is then 
defined as the Laplace transform of the susceptibility
\begin{equation}
\dfrac{\ep(\x,z)}{\ep_\circ} = 1 + \dfrac{1}{2\pi} 
\displaystyle\int_0^{\infty} \exp[i z \t] \chi(\x,\t) \, d \t \, . 
\label{def-ep}
\end{equation}
Since $t$ is positive in the integral above, this permittivity is well-defined for 
complex frequency $z = \om + i \eta$ with positive 
imaginary part Im$(z) = \eta >0$. Moreover, its derivative with respect to the complex 
frequency remains well defined since the function in the integral 
\begin{equation}
\dfrac{d \ep}{d z}(\x, z) = \dfrac{\ep_\circ}{2\pi} 
\displaystyle\int_0^{\infty} (it) \exp[i z \t] \chi(\x,\t) \,  d \t \, 
\label{analytic-ep}
\end{equation}
has exponential decay for Im$(z)>0$. It follows that the permittivity $\ep(\x,z)$ 
is an analytic function in the upper half plane of complex frequencies $z$. Finally, 
it is well-known that, in passive media, the electromagnetic energy must decrease with 
time, and thus the permittivity must have positive imaginary part \cite{Landau8}. 
In particular, the function 
\begin{equation}
\sigma(\x,\om) = \displaystyle\frac{\om \, \text{Im} \, \ep(\x,\om) }{\pi \ep_\circ} \ge 0 \, 
\label{sigma}
\end{equation}
takes positive values. 

In order to define rigorously the inverse of the Helmholtz operator, the auxiliary field 
formalism \cite{Tip97,Tip98,GT10} is used. It is based on the introduction of a new field 
$\A(\t)$ denominated \textit{auxiliary field} which is added to the electromagnetic field 
to form the total vector field $F(\t) = (\E(\t), \B(\t), \A(t)) $ [the only time-dependence 
of the total field $F(\t)$ appears, the other dependences are omitted]. Then, it can be 
shown that Maxwell's equations can be written as the unitary time-evolution equation 
$\dt \F(\t) = - i \mathsf{K} \F(\t)$ where $\mathsf{K}$ is a time-independent selfadjoint 
operator \cite{GT10}. Next, a Laplace transform like (\ref{def-ep}) is applied to this 
time-evolution equation to turn to the complex frequency domain. 
%which becomes $ [z - \mathsf{K} ] \LF(z) = i F(0)$. 
%
%After a Laplace transform, defined for complex numbers $z$ with positive imaginary part by
%\begin{equation}
%\LF(z) = \dfrac{1}{2\pi} \displaystyle\int_\circ^\infty d\t \exp[i z t] \F(\t) \, , 
%\label{Laplace}
%\end{equation}
%the evolution equation (\ref{dtF}) becomes $i z \LF(z) + \F(0) = i \mathsf{K} \LF(z)$ or, 
%equivalently,
%\begin{equation}
%\LF(z) = \dfrac{1}{z - \mathsf{K}} \, i F(0) \, .
%\label{LdtF}
%\end{equation}
Since the operator $\mathsf{K}$ is selfadjoint, the inverse $[z - \mathsf{K}]^{-1}$ is well-defined for 
all complex number $z$ with Im$(z)>0$, and is moreover an analytic function of $z$. Finally, the 
inverse of the Helmholtz operator is obtained by projecting the total fields $\LF(z)$ on 
the electric fields $\LE(z)$ in the equation involving the inverse $[z - \mathsf{K}]^{-1}$ \cite{GM12}. 
Let $\mathsf{P}$ be the projector defined by $\mathsf{P} \LF(z) = \LE(z)$.
%\begin{equation}
%\mathsf{P}_e = \left[ \begin{array}{lcr}
%\mathsf{I}_3 & 0_3 & 0_3 \\ 0_3 & 0_3 & 0_3 \\ 0_3 & 0_3 & 0_3
%\end{array} \right] ,
%\end{equation}
%where $\mathsf{I}_3$ is the $3 \times 3$ unit matrix and $0_3$ is the $3 \times 3$ null 
%matrix. 
Then, the inverse $\inv(z)$ of the Helmholtz operator 
is defined by \cite{GM12}
\begin{equation}
\inv(z) = \mathsf{P} \, \dfrac{1}{z - \mathsf{K}} \, \mathsf{P} \, . 
\label{inv-Helm}
\end{equation}
It can be checked that, from a rigorous calculation based on the Feshbach 
projection formula \cite{Tip00,GT10}, the operator $\inv(z)$ is precisely 
the inverse of the Helmholtz operator defined by equation (\ref{Helm1}).
Since the projector $\mathsf{P}$ is $z$-independent, the inverse 
of the Helmholtz operator has the same analytic properties than the inverse 
$[z - \mathsf{K}]^{-1}$. 

%--------------------------------------------------------------------------------------------------
\section{General properties of reflection and transmission coefficients}
%--------------------------------------------------------------------------------------------------

The most general properties of reflection and transmission coefficients are deduced 
from those of the inverse Helmholtz operator $\inv(z)$ introduced in the previous 
section (\ref{inv-Helm}). 
Indeed, by definition, the inverse $\inv(z)$ is related to the Green's function 
$\mathsf{G}(\x,\y;z)$ by
\begin{equation}
[\inv(z) f ] (\x) = \int_{\mathbb{R}^3} \mathsf{G}(\x,\y;z) f (\y) \, d\y \, , 
\label{fdGreen}
\end{equation}
where $f(\y)$ is an ``admissible'' function. For example, functions $f (\y)$ and $h(\x)$ can be 
chosen arbitrary close to Dirac ``functions'' $\delta(\y -\y_0)$ and $\delta(\x -\x_0)$ centered 
at $\y_0$ and $\x_0$: it implies that 
\begin{widetext}
\begin{equation}
\begin{array}{ll}
\displaystyle\int_{\mathbb{R}^3} \overline{h (\x)} \, [\inv(z) f ] (\x) \, d\x \, 
%& = \displaystyle\int_{\mathbb{R}^3} d\x f (\x) 
%\displaystyle\int_{\mathbb{R}^3} d\y \, \mathsf{G}(\x,\y;z) \, f (\y) \\[4mm]
& = \displaystyle\int_{\mathbb{R}^3} \delta(\x -\x_0) \left[
\displaystyle\int_{\mathbb{R}^3} \mathsf{G}(\x,\y;z) \, \delta(\y -\y_0) \, d\y \right] \, d\x \, = 
\displaystyle\int_{\mathbb{R}^3} \delta(\x -\x_0) \, \mathsf{G}(\x,\y_0;z) \, d\x\\[4mm]
& = \mathsf{G}(\x_0,\y_0;z) \, ,
\end{array}
\label{fdGreen}
\end{equation}
\end{widetext}
which shows that properties of the inverse $\inv(z)$ are directly transposable to the Green's 
function $\mathsf{G}(\x_0,\y_0;z)$. In the particular case of multilayered stacks, 
the Fourier decomposition (\ref{Fx1x2}) is applied. The partial derivatives $\partial / \partial \xs_1$ 
and $\partial / \partial \xs_2$ in Maxwell's equations are replaced by $- i \ks_1$ and $- i \ks_2$, 
and the Fourier transformed $\mathsf{G}(\x_0,\y_0;z)$ is denoted by 
$\widehat{\mathsf{G}}(\kp,\xs_0,\ys_0;z)$, where $\xs_0$ and $\ys_0$ are the $\xs_3$-component of 
the vectors $\x_0$ and $\y_0$. From the expression of the Green's function of 
multilayers \cite{Tom95}, and choosing  $\xs_0$ and $\ys_0$ at the top ($\xs_3 = \xs_u$) or bottom 
($\xs_3 = \xs_d$) interfaces delimiting the multilayer, it can be deduced that
\begin{equation}
\begin{array}{ll}
\widehat{\mathsf{G}}(\kp,\xs_u,\xs_u;z) \!\!
& = - \, \dfrac{z}{2 i \beta_\circ(\kp,z)} \, [1 + r(\kp,z)] \, , \\[4mm]
\widehat{\mathsf{G}}(\kp,\xs_u,\xs_d;z) \!\!
& = - \, \dfrac{z}{2 i \beta_\circ(\kp,z)} \, t(\kp,z) \, ,
\end{array}
\label{fdGreenRT}
\end{equation}
%\begin{equation}
%\widehat{\mathsf{G}}(\kp,\xs_u,\xs_u;z) = - \, \dfrac{z}{2 i \beta_\circ(\kp,z)} \, [1 + r(\kp,z)] \, , 
%\label{fdGreenRT}
%\end{equation}
where $\beta_\circ(\kp,z)$ is uniquely defined as the square root 
$\beta_\circ(\kp,z) = \sqrt{z^2\ep_\circ\mu_\circ - \kp^2}$ with positive imaginary part. Note that 
$\beta_\circ(\kp,z)$ is also analytic for $z$ with positive imaginary part. 

We are now ready to derive the general properties of the reflection and transmission coefficients. 
The analytic properties are deduced directly from relations (\ref{fdGreenRT}) and (\ref{fdGreen}). 
The analytic property of inverse Helmholtz operator $\inv(z)$ implies the well-know result \cite{GO91,TBP97} 
stating that the coefficients $r(\kp,z)$ and $t(\kp,z)$ are analytic functions 
in the half plane of complex frequencies $z$ with positive imaginary part. We propose herein to extend 
this property to the complex wave vector $\kp$ associated with the two-dimensionnal invariance of the 
system. After the Fourier decomposition (\ref{Fx1x2}), the partial derivatives $\partial / \partial \xs_1$ 
and $\partial / \partial \xs_2$ in Maxwell's equations are replaced by $- i \ks_1$ and $- i \ks_2$. 
The resulting Fourier transformed operator $\widehat{\mathsf{K}}(\kp)$ is selfadjoint for real $\ks_1$ 
and $\ks_2$. Its definition can be extended to complex $\kp$ (i.e. the components $\kappa_1$ and 
$\kappa_2$ are complex numbers) and, in that case, $\widehat{\mathsf{K}}(\kp)$ is no more selfadjoint. 
Then, it can be shown that the imaginary part of the operator $[z-\widehat{\mathsf{K}}(\kp)]$ defined as 
\begin{equation}
\text{Im}[z-\widehat{\mathsf{K}}(\kp)] = \dfrac{1}{2i} \Big\{ [z-\widehat{\mathsf{K}}(\kp)] - 
[z-\widehat{\mathsf{K}}(\kp)]^\dagger \Big\} \, , 
\label{ImK}
\end{equation}
only depends on the imaginary parts of $z$ and $\kp$. Simple calculations show that this imaginary part 
above is semi-bounded:
\begin{equation}
\text{Im}[z-\widehat{\mathsf{K}}(\kp)] \ge \text{Im}(z) - c \, |\text{Im}(\kp)| \, , 
\label{ImKbounded}
\end{equation}
where $|\text{Im}(\kp)| = \sqrt{\text{Im}(\ks_1)^2+\text{Im}(\ks_2)^2}$ and $c = 1/\sqrt{\ep_\circ \mu_\circ}$.
It follows that this imaginary part remains strictely positive if $\text{Im}(z) > c \, |\text{Im}(\kp)|$.
Under this condition, the corresponding inverse $[z-\widehat{\mathsf{K}}(\kp)]^{-1}$ is analytic with respect to 
all the complex variables $z$ and $\kp$, which yields the following result. 

{\textbf{Result 1.}} {\textit{The reflection and transmission coefficients, $r(\kp,z)$ and $t(\kp,z)$,
are analytic functions of the complex variables $z$ and $(\ks_1,\ks_2) = \kp$ in the domain
defined by $\text{Im}(z) > c \, |\text{Im}(\kp)|$.}}

Next, a new property is derived from an analogy with the passivity requirement for the electric 
permittivity. It is well-known that the passivity requires for the function $\ep(\x,z)$ to 
have its imaginary part to be positive, see equation (\ref{sigma}). Using a generalization \cite{GT10} 
of the Kramers and Kronig relations, it can be shown that
\begin{equation}
\text{Im} \{ z \ep(\x,z) \} \ge \text{Im} \{ z \ep_\circ \} > 0 \, . 
\label{Im_ep}
\end{equation}
A similar relationship can be established for the inverse of the Helmholtz operator from
\begin{equation}
\begin{array}{ll}
- \text{Im} \dfrac{1}{z - \mathsf{K}} & = \dfrac{1}{2i} \left[ 
\dfrac{1}{\overline{z} - \mathsf{K}} - \dfrac{1}{{z} - \mathsf{K}} \right] \\[4mm]
& = \dfrac{1}{z - \mathsf{K}} \, \text{Im}(z) \, 
\dfrac{1}{\overline{z} - \mathsf{K}} > 0 \, .
\end{array}
\label{Im_res}
\end{equation}
Again, this expression can be extended to complex values of the wave vector $\kp$. According to the 
relation (\ref{ImKbounded}), if $\text{Im}(z) > c \, |\text{Im}(\kp)|$, then the 
inverse $[z-\widehat{\mathsf{K}}(\kp)]^{-1}$ is well defined and has positive imaginary part 
since 
\begin{equation}
\begin{array}{l}
- \text{Im} \dfrac{1}{z - \widehat{\mathsf{K}}(\kp)} \ge \\[4mm]
\quad \quad \dfrac{1}{z - \widehat{\mathsf{K}}(\kp)} \, 
\big[ \text{Im}(z) - c \, |\text{Im}(\kp)| \, \big] \,
\dfrac{1}{\overline{z} - \widehat{\mathsf{K}}(\overline{\kp})} > 0 \, .
\end{array}
\label{Im_Fou_res}
\end{equation}
It has to be noticed that, in the case of this ``passivity'' property, the extension to complex wave 
vector $\kp$ is of vital importance for the transposition to the coefficients 
$r(\kp,z)$ and $t(\kp,z)$ at non normal incidence $\kp \neq \boldsymbol{0}$. Indeed, for all propagative 
waves it is possible to choose $\kp = z \u $ with $\u^2 < \sqrt{\ep_\circ \mu_\circ}$, so that the relation 
$\text{Im}(z) > c \, |\text{Im}(\kp)|$ remains true. Hence, the square root $\beta_\circ(\kp,z)$ 
in equation (\ref{fdGreenRT}) can be written $\beta_\circ(\kp,z) = z  \sqrt{\ep_\circ \mu_\circ - \u^2}$, 
and the following result can deduced from (\ref{Im_Fou_res}) and the first line of 
(\ref{fdGreenRT}).

{\textbf{Result 2.}} The reflection coefficient $r(\kp,z)$ must satisfy
\begin{equation}
\text{Re} \{1 + r(z \u, z) \} > 0 \, , %\quad \text{Re} \{ t(z \u, z) \} > 0 \, , 
\label{Re_r}
\end{equation}
{\textit{where $\u$ is defined by $\kp = z \u$ and has modulus square $\u^2 <  \ep_\circ \mu_\circ$.}}

In practice, the function $\{1 + r(z \u, z) \} $ %and  $\{ t(z \u, z) \}$ 
has positive real part 
for all complex frequency $z$ with positive imaginary part, and for all fixed angle $\theta$ from 
normal incidence defined by $\u^2 = \ep_\circ \mu_\circ \sin^2 \theta $. Thus the result (\ref{Re_r}) 
can be applied to only propagating waves. For evanescent waves, if the wave vector is written 
$\kp = z \u$ with $\u^2 > \ep_\circ \mu_\circ$, both relations (\ref{Im_res}) and (\ref{Im_Fou_res}) 
need not to be correct, and then there is no simple condition for the reflection coefficient. 

Finally, an addionnal result can be obtained evaluating the Green's function at a distance $\ds/2$ 
above the top interface delimiting the multilayered stack: $\xs_0 = \ys_0 = \xs_u + \ds/2$. 
This distance introduces the phase shift $\exp[i\beta_\circ(\kp,z)\ds]$ and the first line 
of (\ref{fdGreenRT}):
\begin{equation}
\begin{array}{l}
\widehat{\mathsf{G}}(\kp,\xs_u+\ds{\resub{/2}},\xs_u+\ds{\resub{/2}};z) = \\[2mm]
\quad \quad- \, \dfrac{z}{2 i \beta_\circ(\kp,z)} \{ 1 + \exp[i\beta_\circ(\kp,z)\ds] r(\kp,z) \} \, . 
\end{array}
\label{fdGreenR}
\end{equation}
Thus the result 2 can be generalized as follows. 

{\textbf{Result 3.}} {\textit{The reflection coefficient $r(\kp,z)$
must satisfy}}
\begin{equation}
\text{Re} \{ 1 + \exp[i\beta_\circ(\kp,z)\ds] r(\kp,z) \} > 0 \, , 
%\quad | 1 + \exp[2i\beta_\circ(\kp,z)\ds] r(\kp,z) | > 0 \, , 
\label{Rnv}
\end{equation}
{\textit{for all complex variables $z$ and $(\ks_1,\ks_2) = \kp$ in the domain
defined by $\text{Im}(z) > c \, |\text{Im}(\kp)|$.}}

It is stressed that the results 2 and 3 extend the well known relation \cite{TBP97} derived from 
the calculation of the Poynting vector flux for real frequency and wavevector. Indeed, 
let $z=\omega$ be purely real, then $\exp[i\beta_\circ(\kp,z)\ds]$ becomes a pure phase shift with 
unit modulus. let $d$ vary, then relation (\ref{Rnv}) shows that $r(\kp,z)$ is located on a circle 
with radius below unity, i.e. $ | r(\kp, \omega) | \leq 1 $.

%--------------------------------------------------------------------------------------------------
\section{Phase retrieval}
%--------------------------------------------------------------------------------------------------

\subsection{Kramers-Kronig relations at a fixed oblique incidence}

In this section, the Kramers-Kronig relations are used to retrieve the phase of reflection and transmission 
coefficients from their modulus (intensity of the field). Let $f(z)$ be an analytic function in the half plane 
of complex numbers $z$ with positive imaginary part, which vanishes at the infinity: $f(z) \longrightarrow 0$ 
for $|z| \longrightarrow \infty$. Then Kramers and Kronig relations leads to 
\begin{equation}
\text{Im} \{ f(\om) \} = \dfrac{1}{\pi} \mathcal{P}\displaystyle\int_{\mathbb{R}} 
\dfrac{\text{Re} \{ f(\nu) \}}{\omega - \nu} \, d \nu \, ,
\label{KKf}
\end{equation}
where the symbol $\mathcal{P}$ means that the Cauchy principal value of the integral. In order to separate 
the phase from the modulus of a function, it is convenient to apply the complex logarithm $\ln$. It
is important to note that, the complex logarithm must be applied to a non vanishing function to preserve 
the analytic properties. This condition will be ensured by the results 2 and 3 
for the function containing the reflection coefficient. 
%%%%%%%%%%%%%%%%%%%%%%%%%% resub
{\resub{As to the transmission coefficient, the well know result on the Poynting 
vector flux can be used: for Im$z>0$
\begin{equation}
1 - | r(\kp, z) |^2 - | t(\kp, z) |^2 > 0 \, .
\end{equation}
It implies that the modulus of the transmission and reflection coefficients 
is strictly smaller than unity. In particular, {\textit{The transmission 
coefficient $t(\kp,z)$ must satisfy}}
\begin{equation}
 | \exp[i\beta_\circ(\kp,z)\ds] t(\kp,z) | < 1 \, , 
\label{Tnv}
\end{equation}
{\textit{for all complex variables $z$ and $(\ks_1,\ks_2) = \kp$ in the domain
defined by $\text{Im}(z) > c \, |\text{Im}(\kp)|$.}}
Also, is well known that the transmission coefficient}} cannot 
vanish \cite{TBP97} [$t(\kp, \omega)=0$ implies that the field must vanish 
in all the space, which is impossible when the multilayered stack is 
illuminated by a plane wave]. 
%%%%%%%%%%%%%%%%%%%%%%%%%% resub
{\resub{Finally,}} it is stressed that the extension of analytic properties to complex 
wave vector, stated in result 1, is crucial to keep fixed 
the angle of incidence for all (complex) frequencies. Thus, for 
$\kp = z \u$, the Kramers and Kronig 
relations (\ref{KKf}) can be applied to functions
\begin{equation}
\begin{array}{l}
f_t(z) = \ln \{ t(z \u, z) \} \, , \\[2mm]
f_r(z) = \ln \{ {\resub{1 + \exp[i\beta_\circ(z\u,z)\ds] r(z \u, z)}} \} \, .
\end{array}
\label{fRT}
\end{equation}
Thus we can propose solution to determine the phase of the transmission and 
reflection coefficients using the relations
\begin{equation}
\begin{array}{l}
\text{Phase} \{ t(\om \u, \om) \} = 
\dfrac{1}{2 \pi} \mathcal{P}\!\!\displaystyle\int_{\mathbb{R}} 
\dfrac{\ln | t(\nu \u, \nu) |^2 }{\omega - \nu} \, d \nu \, , \\[4mm]
\text{Phase} \{ {\resub{1+\exp[i\beta_\circ(\om\u,\om)\ds] 
r(\om \u, \om)}} \} = \\[2mm]
\quad\quad \dfrac{1}{2 \pi} \mathcal{P}\!\!\displaystyle\int_{\mathbb{R}} 
\dfrac{\ln | {\resub{1+\exp[i\beta_\circ(\nu\u,\nu)\ds] r(\nu \u, \nu)}} |^2 }
{\omega - \nu} \, d \nu \, .
\end{array}
\label{KKRT}
\end{equation}
As a final step, the modulus and the phase of the reflection coefficient 
$r(\om \u, \om)$ 
is directly deduced from the knowledge of both 
the modulus and the phase of 
$\{ \, {\resub{1+\exp[i\beta_\circ(\om\u,\om)\ds] r(\om \u, \om)}} \}$.

It is stressed that our result, with the wavevector $\kp=z\u$ related to the frequency $z$ to keep 
constant incidence angle, is more general than one with the wavevector $\kp$ set to a real constant. 
Indeed, in the case where $\kp$ is constant, there is no need to extend the 
properties stated for the frequency to complex wavevector. Also, with $\kp$ constant, the Kramers-Kronig 
relations are difficult to exploit since the incident angle varies according to the frequency, and 
especially both regimes of propagating and evanescent waves are addressed when the frequency describe 
the whole spectrum. 

In practice, the intensity is obtained in a finite interval of frequencies while 
the use of Kramers-Kronig relations requires measurements over all the frequency spectrum. This 
difficulty, which is not considered in this paper, can be overcome by a normalization procedure 
as proposed in \cite{GO91}.

In the next subsection, solutions are discussed to measure the modulus (intensity) 
of the functions $| t(\om \u, \om) |^2 $ and 
$| {\resub{1+\exp[i\beta_\circ(\om\u,\om)\ds] r(\om \u, \om)}} |^2 $
from which the phase of reflection and transmission coefficients can be deduced. 

\subsection{Discussion}

First, it is stressed that our results provide a rigorous proof showing 
that the techniques already established \cite{GO91} to retrieve the 
phase from the intensity can 
be used in non-normal incidence. For instance formula (\ref{KKRT}) can be used directly for the 
transmission  coefficient to retrieve the phase of the transmission coefficient from the 
measurement of the transmitted intensity. %(see figure \ref{fig2}, left panel). 

\begin{figure}
\centerline{\includegraphics[width=.9\columnwidth]{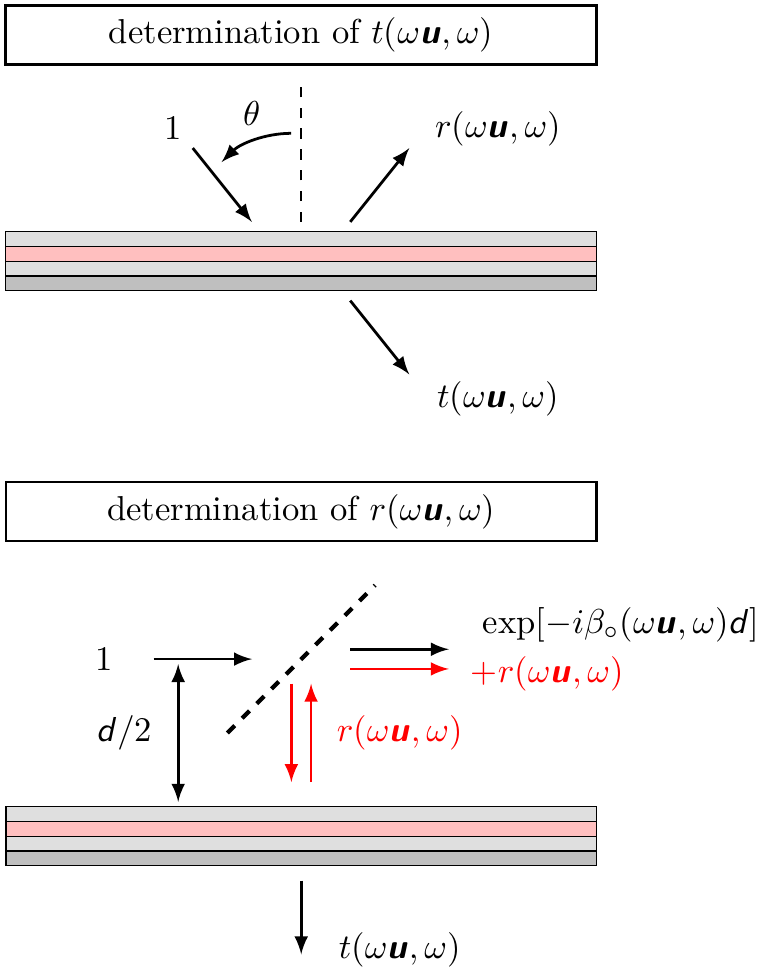}}
\caption{Top panel: Direct determination of the transmitted intensity 
$|t(\om\u,\om)|^2$. 
Lower panel: Determination of the quantity 
$| {\resub{1+\exp[i\beta_\circ(\om\u,\om)\ds]\,r(\om \u, \om)}} |^2$ 
from which the phase of the reflection coefficient $r(\om \u, \om)$ can be deduced.\label{fig2}}
\end{figure}

As to the reflection coefficient, formula (\ref{KKRT}) invites us to consider the
quantity $|1 + r(\kp, \om)|^2$ which is the field intensity at the top interface 
delimiting the multilayered stack. This quantity can be determined by measuring 
several physical quantities. For instance, the fluorescence of excited atoms (or molecules) 
located at the top interface can be directly related to the the desired field intensity 
using the fluctuation-dissipation theorem \cite{Rytov89,Joulain09}.
To obtain this effect, it is necessary to add at the top interface of the multilayer 
a very thin absorbing layer in which the emitters are implanted. Note that the additional 
layer has to be sufficiently thin to avoid the significant perturbation of the 
reflectivity properties. 
Also, if some roughtness is added to the top interface, then the scattered intensity 
is proportionnal to the field intensity. In such case, the knowledge of the factor 
of proportionality requires a precise knowledge of the structure. It is stressed that 
this can be probably more simple to obtain such factor than the whole set of complex zeros 
of the reflection coefficient and the associated Blaschke factors. 
Finally, it has to be noticed that %, with an interferometer, 
the reflected field can be superimposed to the incident field in order to obtain directly 
the quantity $| {\resub{1+\exp[i\beta_\circ(\om\u,\om)\ds]\, r(\om \u, \om)}} |^2 $ 
(see figure \ref{fig2}, lower panel). Here, the convenience to use a setup like the one shown 
on figure (\ref{fig2}) to obtain 
$| {\resub{1+\exp[-i\beta_\circ(\om\u,\om)\ds]\, r(\om \u, \om)}} |^2$ has to be compared with the direct 
measurement of the phase, in order to estimate the relevance of such a proposition. 

%--------------------------------------------------------------------------------------------------
\section{Conclusion}
%--------------------------------------------------------------------------------------------------

We went further in the investigation of causality principles in the sense that some multilayer 
responses were shown to be analytic when both Fourier variables ($\omega$ and $\kp$) are extended 
to the domain of the complex plane defined by Im$(z) > c |$Im$(\kp)|$. This result allowed us 
to extend the Kramers-Kronig relationships to the more general situation of oblique incidence. 
These relations can be applied to the complex logarithm of the energy transmission function, but 
also to the quantity 
$\ln |{\resub{1+\exp[i\beta_\circ(\om\u,\om)\ds] \,r(\om\u, \om)}}|^2$, 
thus providing a 
new way to investigate phase. From the point of view of experiment, this last quantity can be 
approach with luminescence, scattering measurements, or using an interferometer. 

\bibliographystyle{osa}
\bibliography{KKR-AO}

\end{document}